# Distinguishing photothermal and photoelectrokinetic contributions to flow in light-activated microchannel architectures

Sudharshan Anandan[1], Yuhang Fang[1], Lily Farmer[2], Maria Irineu[1], Shanmukhi Sripada[1], Tyler Hughes [1], Jared Nathan[1], Sathish Rajendran[1], Steven Wereley[1], Amy Marconnet[1], Thomas Beechem[1], Jeffrey L. Moran[3,4*], David M. Warsinger[1*]

[1]School of Mechanical Engineering, Purdue University, West Lafayette, IN.

[2]Davidson School of Chemical Engineering, Purdue University, West Lafayette, IN

[3]Department of Mechanical Engineering, George Mason University, Manassas, VA

[4]Department of Bioengineering, George Mason University, Manassas, VA

*Corresponding authors: dwarsing@purdue.edu; jmoran23@gmu.edu

David M. Warsinger

Associate Professor

School of Mechanical Engineering, Purdue University

585 Purdue Mall, West Lafayette, IN 47907-2088

Offices: BRK 2266

Jeffrey L. Moran

Associate Professor and Associate Chair for Graduate Studies

Department of Mechanical Engineering, George Mason University

Phone: 703-993-5636

Campus: Science and Technology

Office Location: Institute for Advanced Biomedical Research, Room 3015

Mailing Address: 10920 George Mason Circle, Mail Stop 4C7, Manassas, VA 20110

# Abstract:

Light-driven flows near Janus surfaces could power microfluidic devices without external hardware. These flows are often attributed to photoinduced electro- or diffusio-osmosis, but illumination also heats the surface, potentially driving natural convection. Here, we map flow and temperature around a UVA-illuminated $TiO_2$–Au Janus membrane using depth-resolved particle tracking and infrared thermometry. Recirculating rolls form on both sides, with no measurable net flow across the membrane. Their cubic, reversing velocity profile identifies horizontal (Hadley) convection, driven by a lateral temperature gradient as the gold film absorbs light. The measured profile agrees with wall-corrected Hadley theory using one fitted parameter. Adding 100 mM NaCl to suppress electrokinetic flows leaves the profile unchanged, indicating that convection dominates. Simulations show that convection falls below 1 $\mu m\ s^{-1}$ in thin channels, where surface-driven osmotic flows could dominate. Thus, device geometry and surface chemistry determine which mechanism governs light-driven microfluidic flows.

# INTRODUCTION

Autonomous and precise microscale fluid control is critical for applications including lab-on-a-chip diagnostics and remote water sensing. Achieving practical setups for microfluidic flow manipulation without external pumps or moving parts could enable lower-cost, simpler devices for point-of-care diagnostics and field-deployable water-quality monitoring applications, but to date remains a significant challenge. Early strategies relied on chemical reactions, notably hydrogen peroxide-driven electrocatalysis with bimetallic nanorods or membranes, to power autonomous pumps[1–3]. However, these methods require reagents like hydrogen peroxide, which are unsuitable for biological samples and impractical for water treatment, motivating the search for alternative, non-toxic approaches.

Light offers a promising, non-contact means of activating microfluidic flows, eliminating the need for chemical additives. Both photoelectrokinetic and photothermal effects have enabled light-controlled micromotors and pumps. In photoelectrokinetic mechanisms, light redistributes ions or charges, generating electrical body forces that drive electrokinetic or diffusio-osmotic flows. In contrast, photothermal effects stem from local temperature gradients created by light absorption, producing buoyancy-driven flows, surface-tension gradients, and thermophoretic motion—each measurable by particle tracking and scaling analysis. Under an applied AC electric field, membrane heating can also generate electrothermal flow via temperature-dependent gradients in fluid permittivity and conductivity[4,5].

For example, two-faced $TiO_2$-Au "Janus" microspheres self-propel in water under ultraviolet (UV) illumination, with velocities that scale with light intensity[6]. Immobilized $TiO_2$-Au structures can also generate fluid flow under UV light, highlighting the promise of light-driven pumping in microfluidics[7]. However, in many systems, unambiguously identifying the underlying mechanism remains challenging. Rigorous assessment demands comprehensive materials characterization, careful control experiments, and precise attention to cell architecture—factors often overlooked in proof-of-concept micropump studies. Device size scale is critical, as flows result from a balance between viscous forces and driving body forces, contributing to inconsistencies in distinguishing photoelectrokinetic from photothermal effects. Thus, systematic integration of materials analysis, flow control, and quantitative measurement is essential to resolve the driving forces and advance light-responsive microfluidic technologies toward practical application.

Here, we address this gap by systematically investigating the mechanisms underlying light-induced flow in a membrane-based system. This multidisciplinary problem requires an integrated framework that combines control experiments, materials characterization, and device design to accurately identify the driving mechanism. Unlike previous studies centered on proof-of-concept demonstrations, we employ a unified approach—incorporating control experiments, optical simulations, non-isothermal flow simulations, and scaling analysis—to disentangle photoelectrokinetic and photothermal effects. Control experiments varying ionic strength and absorption calculations show that photothermal convection, not photoelectrokinetic flow, is the primary driver of fluid motion under UV illumination. Depth-resolved velocity measurements reveal strong, bidirectional horizontal convection below the membrane, while simulations demonstrate a similar convection roll above it—a feature largely overlooked in previous studies. Optical transfer matrix calculations, non-isothermal flow simulations, and scaling analysis corroborate these findings, emphasizing the importance of comprehensive mechanistic validation for the advancement of light-powered microfluidic technologies.

## RESULTS

### Membrane architecture and experimental setup

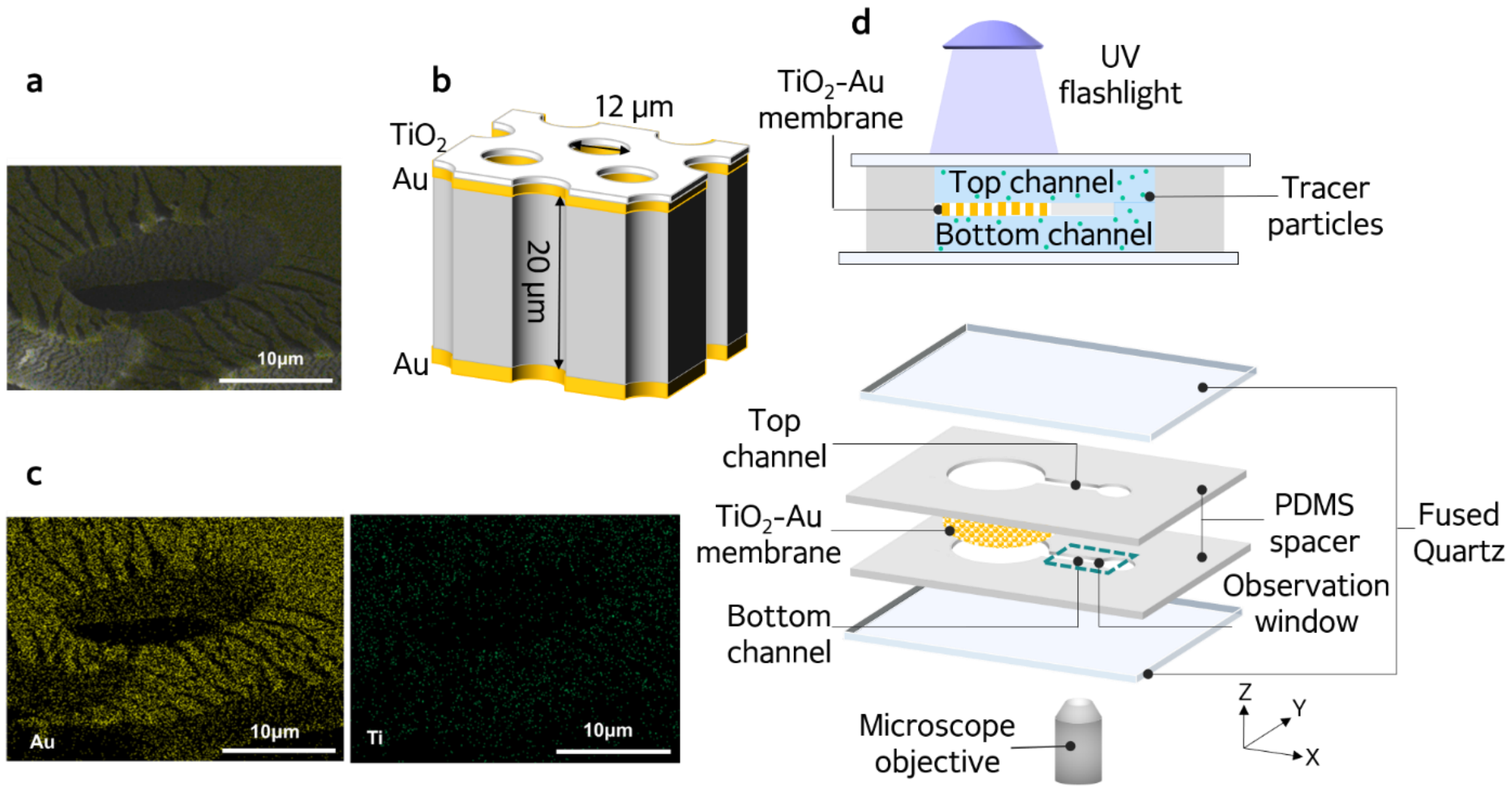

Figure 1. | **Illustration of the membrane architecture and exploded view of the experimental setup** (a) Scanning electron microscopy image of a single pore coated with $TiO_2$-Au (b) schematic of the membrane architecture (c) EDX images of the conformal coatings of Au and $TiO_2$ on the membrane (d) Exploded-view schematic of the measurement chamber: UV flashlight illumination, quartz windows, PDMS spacer, $TiO_2$-Au membrane, liquid chamber, and inverted-microscope objective for depth-resolved particle tracking (coordinate axes defined)

Figure 1 illustrates the two-layer PDMS fluidic device engineered to visualize light-activated flows. The device consists of two laser-cut PDMS channels with asymmetric dumbbell-shaped profiles, each bearing two circular wells (diameters: 19 mm and 6 mm) connected by an observation channel (1 mm width, 11 mm length). Heating is localized by restricting the illuminated, coated surface to the membrane over the large well. The observation channel is blocked with a 0.1 mm double-sided acrylic adhesive tape, separating the top and bottom PDMS channels along the observation channel, so that the two layers connect only through the membrane and the cold through-well. The fluidic assembly is vertically stacked and symmetric: the bottom layer is a polished quartz slide supporting a thick PDMS cavity, sealed with the double-sided acrylic adhesive tape to seal the primary channel while leaving the terminal wells exposed. The track-etched polycarbonate membrane (pore diameter and porosity detailed in Methods), sputter-coated with Au on both sides and with $TiO_2$ on the top side, is aligned over the larger well. The assembly is completed with a symmetric top PDMS cavity and a top quartz slide. Each layer comprises a large cylindrical reservoir (hot well) connected to a smaller cylindrical reservoir (cold through-well) via the observation channel, and the total internal vertical fluid column extends through the assembled stack, enabling controlled studies of flow on both sides of the membrane. The membrane is irradiated by a localized UV LED source focused on the large well. Flow fields are quantified by tracking fluorescent polystyrene tracer particles (TrackPy) at known focal planes. Depth-resolved scans are performed along the central axis of the observation channel, enabling direct visualization and quantification of flow profiles in response to optical activation.

## Photothermal convection in the flow cell

Under dark conditions, negligible flow velocities were detected in the observation channel. When the TiO2-coated membrane surface was irradiated with UV light, an antisymmetric cubic flow profile emerged (Fig. 2). The same profile, within experimental uncertainty, was observed in 100 mM NaCl (Fig. 2b). This result contradicts predictions of photoelectrokinetic mechanisms proposed in earlier studies[7,8] for this geometry, which suggest that increasing ionic strength reduces flow speed significantly. Because photoelectrokinetic slip scales with the zeta potential and the electric field across the electric double layer, reducing the Debye length by more than an order of magnitude (to ≈1 nm at 100 mM NaCl) should strongly suppress the slip. However, the profile shape remains unchanged. Measurements focus on the channel below the membrane.

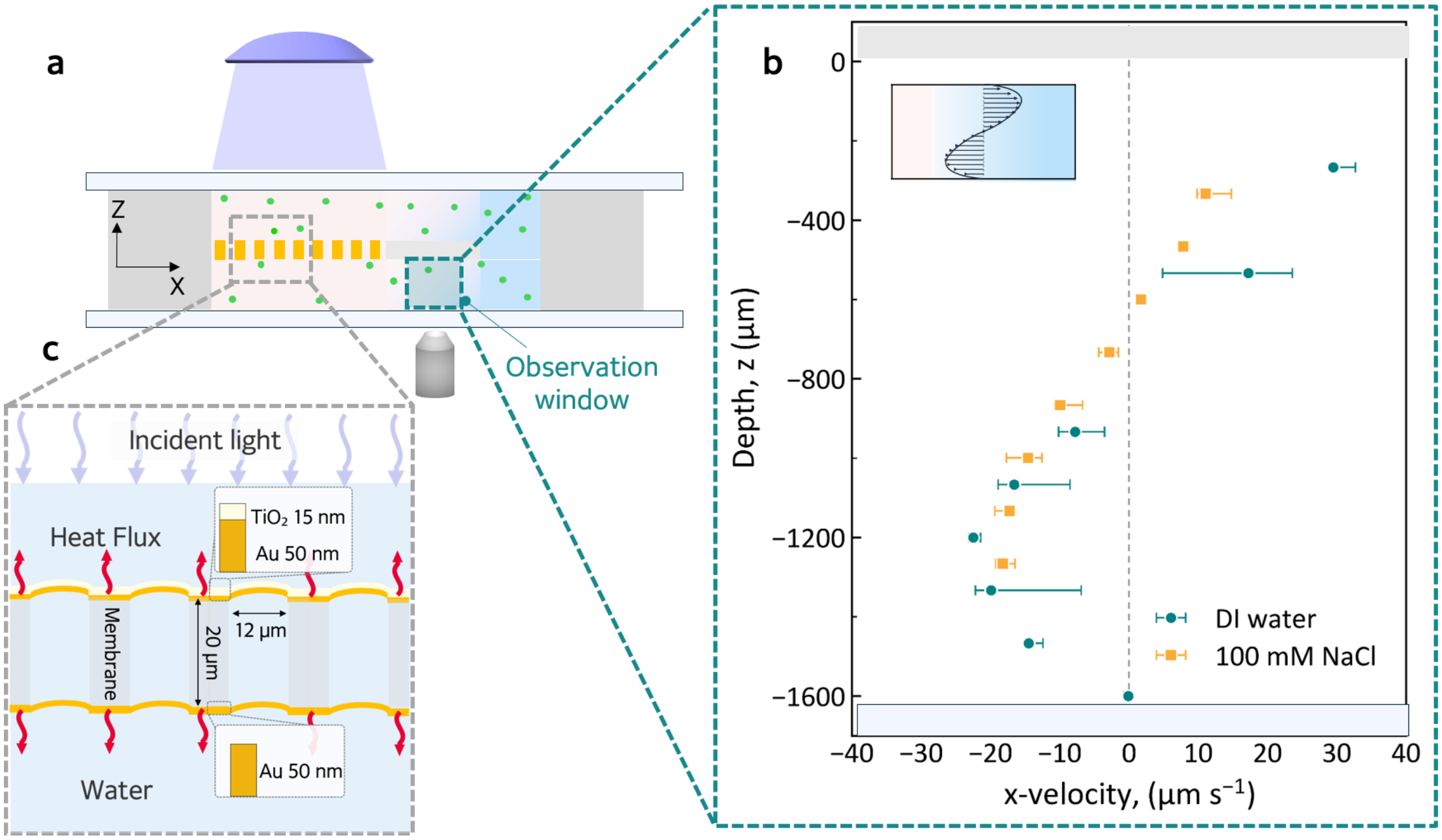

**Figure 2. | Depth-resolved velocimetry distinguishes photothermal from photoelectrokinetic flows in an illuminated Au/$TiO_2$ Janus membrane channel**. (a) Schematic x–z cross-section of the microchannel device under UV illumination. The channel has transparent top and bottom walls and is bisected by a $TiO_2$–Au-coated membrane. Green tracer particles in the fluid are imaged for particle-tracking velocimetry. The dashed box marks the region magnified in (c). (b) Horizontal velocity ($v_x$) versus depth (z), measured near the center of the observation channel (about 4–6 mm from the hot-well edge) in DI water and 100 mM NaCl (symbols as in the legend). Horizontal bars indicate bootstrapped 95% confidence intervals. The reversal of $v_x$ near mid-depth creates an S-shaped profile characteristic of photothermal (Hadley-type) convection. (c) Magnified view of the region shown in (a), illustrating the membrane coating (Au, 50 nm; $TiO_2$, 15 nm) and pore dimensions (depth, 20 µm; diameter, 12 µm). Absorption by the Au layer heats the fluid, creating a lateral temperature gradient that drives buoyant recirculation (red arrows).

The upper channel could be imaged only partially because of the travel limits of the microscope stage, but it showed a qualitatively similar recirculating pattern, consistent with the simulations (Fig. 4a). The observed antisymmetric flow profile is attributed to a lateral temperature gradient along the x-direction of the observation channel. Specifically, UVA light absorption and concomitant photothermal heating in the underlying gold film results in localized heating of the fluid above and below it. Given the minimal membrane thickness, thermal conduction from the gold film to the fluid generates a lateral temperature gradient across both the top and bottom wells adjacent to the membrane. Consequently, a hot well is established on the left and a cold well on the right of the observation channel both above and below the membrane. This thermal gradient induces a Hadley-type convection cell, characterized by flow at the bottom of the observation channel moving from the cold side to the hot side (right to left), while mass conservation necessitates a counterflow at the top of the channel, thereby producing the observed antisymmetric flow profile.

## Thermometry and thermal mapping

Light absorption induces photothermal heating in the gold film, which creates a lateral temperature gradient between the hot and cold wells. To determine where light converts to heat, we used a transfer-matrix model to calculate absorption within the water/$TiO_2$/Au/polycarbonate stack at the LED peak wavelength (369 nm; Methods). The gold layer absorbs about 87% of the incident light, because the photon energy (3.4 eV) exceeds the interband absorption threshold of gold (about 2.4 eV[9]). The $TiO_2$ layer absorbs only about 1%, leaving little light to drive photochemistry there; the membrane therefore acts primarily as a heater above the hot well. Infrared imaging of the cell on a temperature-controlled stage corroborates this finding (Fig. 3c). Before illumination, the temperature was uniform at 22.0 °C. After 10 minutes of illumination, the temperature decreased smoothly from approximately 27 °C at the hot-well center (P1) to 23 °C at the cold-well center (P2) over 23.5 mm (Fig. 3d). Thermal simulations show the temperature difference across the 1 mm quartz slide is at most 0.25 K, so the slide does not significantly bias the infrared estimate of the upper-water-layer temperature. Thermocouple measurements taken immediately afterward in the two wells were 25.4 °C and 21.9 °C, respectively; because the infrared values are 1–2 °C higher, the image describes the profile shape, while the thermocouples provide absolute temperatures. Two-dimensional simulations, using the absorbed flux as a heat source on the membrane, the measured stage and air temperatures, and estimated heat-transfer coefficients as boundaries, reproduce the measured profile without any fitted parameters (Fig. 3b, d).When both quartz slides were replaced with sapphire, which has 25 times higher thermal conductivity, the well-to-well temperature difference dropped to about 0.2 °C (Supplementary Note 5), demonstrating that the gradient is determined by conduction through the cell walls rather than solely by the membrane.

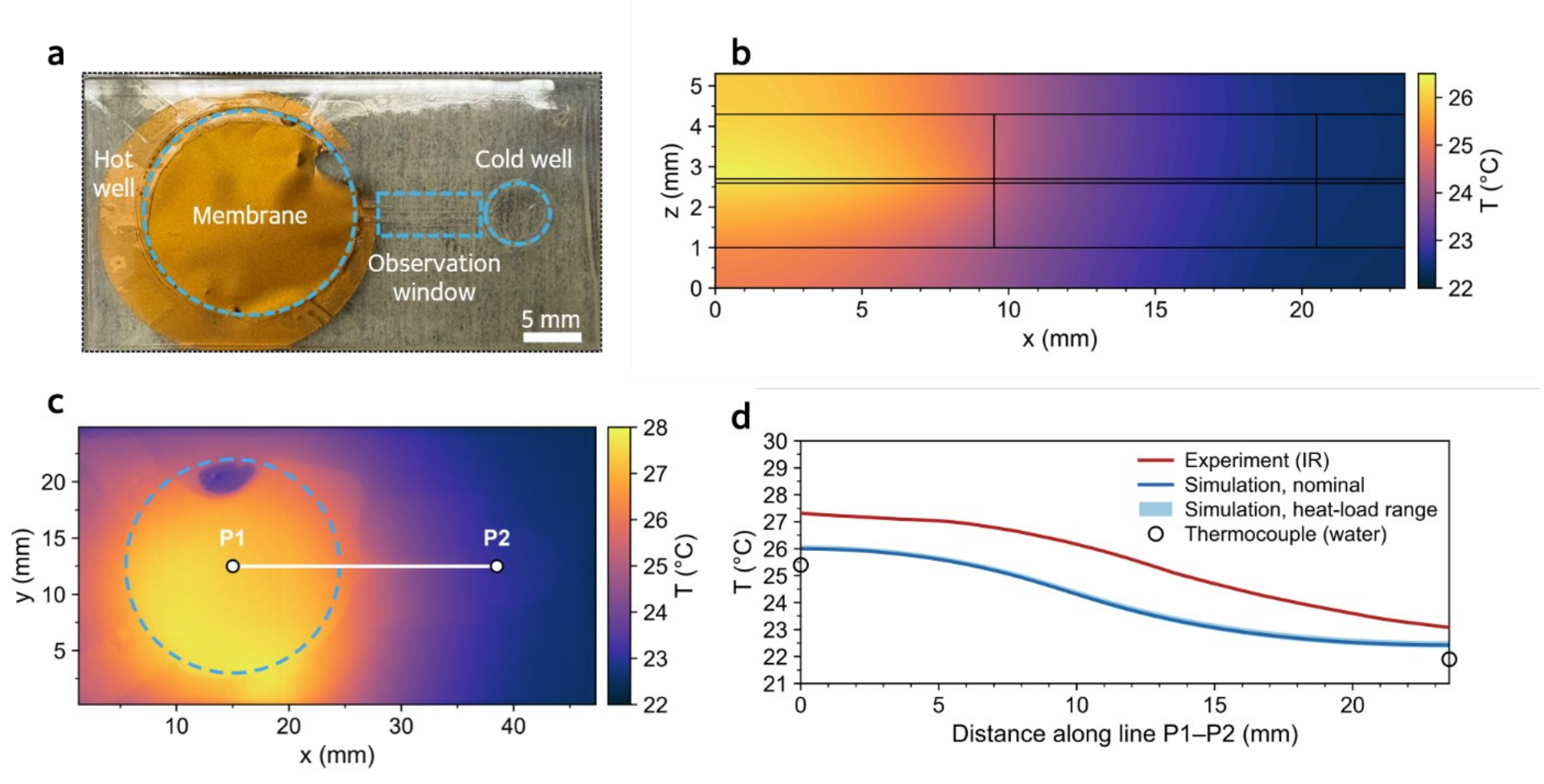

**Figure 3. | Photothermal heating in the gold film generates a lateral temperature gradient.** (a) Photograph of the assembled cell, showing the illuminated membrane (dashed circle) in the hot well, the observation window and the cold well. Scale bar, 5 mm. (b) Simulated steady-state temperature field in the x–z plane along P1–P2 for the infrared imaging conditions: stage at 22 °C, coupled to the cell bottom ($z = 0$) through a contact conductance $h_c = 100$ W m$^{-2}$ K$^{-1}$, and absorbed heat flux of 510 W m$^{-2}$. Lines indicate the model-domain boundaries. (c) Mid-wave infrared image of the cell top after 10 min of illumination. The dashed circle marks the membrane; P1 and P2 mark the centres of the hot and cold wells, respectively. Panels b and c use separate color scales. (d) Temperature profile along P1–P2 (23.5 mm), showing the infrared measurement (red line), thermocouple readings in the upper water layer (open circles) and simulations (nominal, blue line; bounding cases, shaded band). The lower-bound case uses an absorbed heat flux of 490 W m$^{-2}$ and $h_{top} = 15$ W m$^{-2}$ K$^{-1}$; the upper-bound case uses 530 W m$^{-2}$ and $h_{top} = 8$ W m$^{-2}$ K$^{-1}$. All simulations use $h_c = 100$ W m$^{-2}$ K$^{-1}$.

## Numerical and analytical validation of Hadley theory

The system flow is characterized by a Hadley roll driven by the lateral temperature gradient. Simulations under the same thermal conditions as the velocity measurements (microscope stage, still air) reveal a single, long, shallow convection roll in each water layer (Fig. 4a). Water heated above the hot well rises and flows along the top of the layer toward the cold well, then returns along the bottom. In these simulations, the membrane was modeled as an impermeable polycarbonate solid, consistent with the measured velocity profiles, which showed no net flow above about 2 μm $s^{-1}$ (Methods).

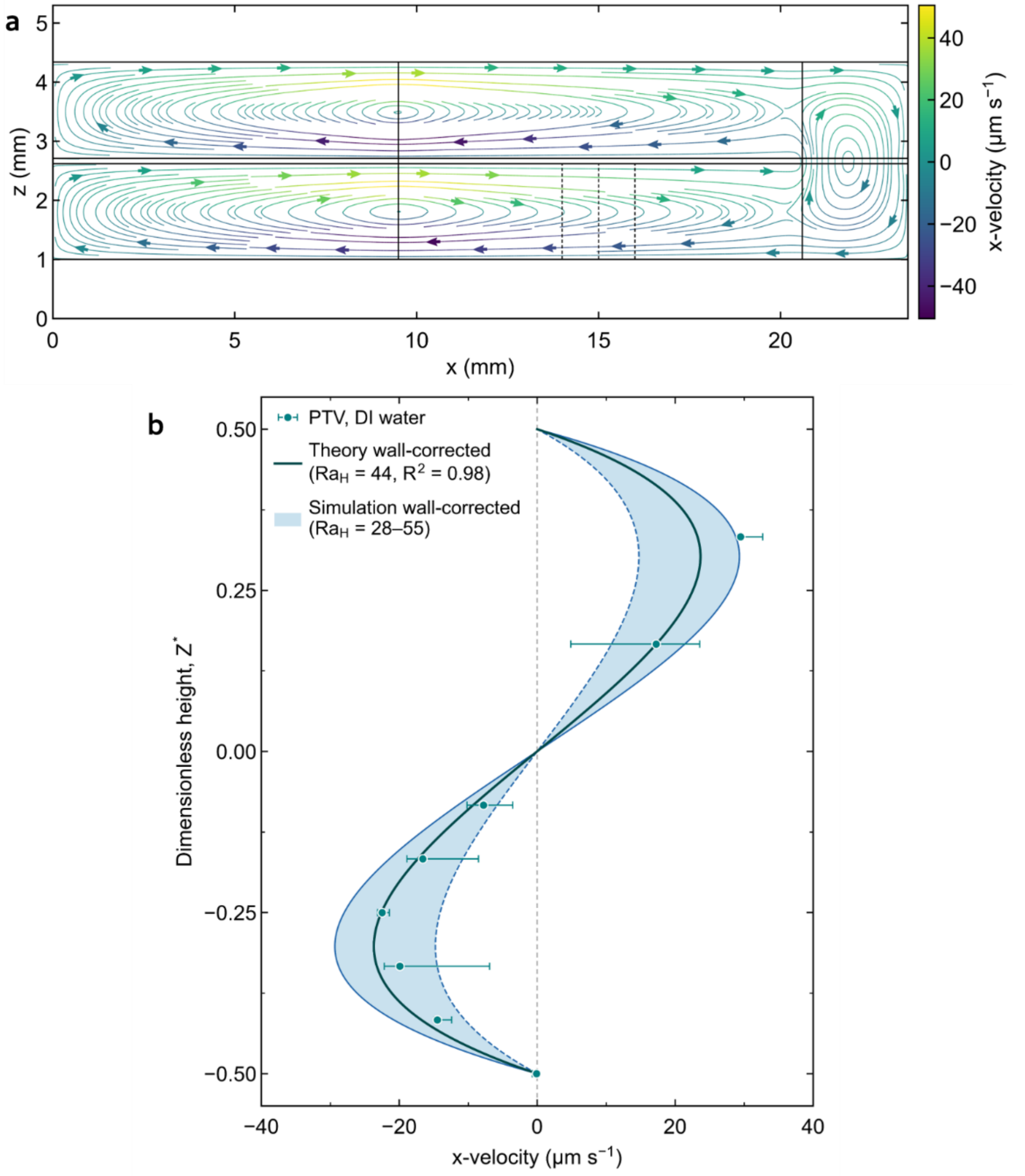

**Figure 4. | Measured velocities align with wall-corrected Hadley convection.** (a) Simulated streamlines in the x–z plane under velocity-measurement conditions are colored by x-velocity, with an impermeable membrane. The two-dimensional simulation lacks side walls, resulting in higher velocities than those observed in (b). Black lines indicate the channel walls, membrane, and the ends of the observation channel (x = 9.5 and 20.5 mm). (b) x-velocity is plotted against the dimensionless height Z* at the bottom channel center. Symbols represent particle tracking in deionized water (medians; bars, bootstrapped 95% confidence intervals). The line shows a one-parameter fit to the wall-corrected Hadley solution ($Ra_H$ = 44, $R^2$ = 0.98), while the shaded band depicts simulations at x = 14–16 mm and for the bounding heat-loss conditions, corrected for side walls ($Ra_H$ = 28–55).

Within the observation channel ($9.5 \leq x \leq 20.5$ mm), streamlines are nearly parallel to the walls, satisfying the condition for Hadley theory. A layer of height H with a horizontal temperature gradient ($\partial T/\partial x$) exhibits a cubic velocity profile[10,11]: forward in the upper half, backward in the lower half. The profile's amplitude is set by the horizontal Rayleigh number ($Ra_H = g\beta|\partial T/\partial x|H^4/(\nu\kappa)$; see Methods). Since $\partial T/\partial x$ varies along the channel, $Ra_H$ is evaluated at the measurement position. The observation channel measures 1 mm wide by 1.6 mm tall. Side walls reduce the flow speed, which we correct for using the four-wall solution, lowering the centerline speed by a factor of 0.73 for this aspect ratio. Fitting this profile to the measured velocities yields $Ra_H = 44$ ($R^2 = 0.98$; Fig. 4b). Simulated profiles, sampled within ±1 mm of the channel center and corrected for side wall effects, span $Ra_H$ = 28–55 and encompass the experimental measurement.

The thermal Péclet number, $Pe = U H/\alpha$, is about 0.3 ($U \approx 26$ μm $s^{-1}$, the measured peak speed in deionized water; H = 1.6 mm, the channel height; $\alpha = 1.46 \times 10^{-7}$ $m^2$ $s^{-1}$, the thermal diffusivity of water at 25 °C) and remains below 0.35 for all ionic strengths, indicating that heat transfer is dominated by conduction, so side walls alter the speed but not the temperature field. Unlike Rayleigh–Bénard convection, Hadley convection has no Rayleigh number threshold for flow onset; thus, any lateral temperature difference will drive flow.

## Ionic screening and channel geometry rule out an electrokinetic origin

Ionic strength does not alter the flow pattern, while channel geometry determines its strength. Photoelectrokinetic flow relies on an electric field acting on the diffuse charge at the wall. Increasing the NaCl concentration from deionized water to 100 mM reduces the Debye layer thickness to about 1 nm and typically suppresses such flow[8]. However, the characteristic reversing cubic velocity profile is observed consistently in deionized water, 10 mM NaCl, and 100 mM NaCl (Fig. 5a; Supplementary Fig. S12). The flow speed remains unchanged from deionized water to 10 mM and only slows in the 100-mM condition. These runs were conducted on separate days, introducing differences in heat loss to the surroundings, yet the velocity profile shape remains consistent (Supplementary Notes 6 and 8).

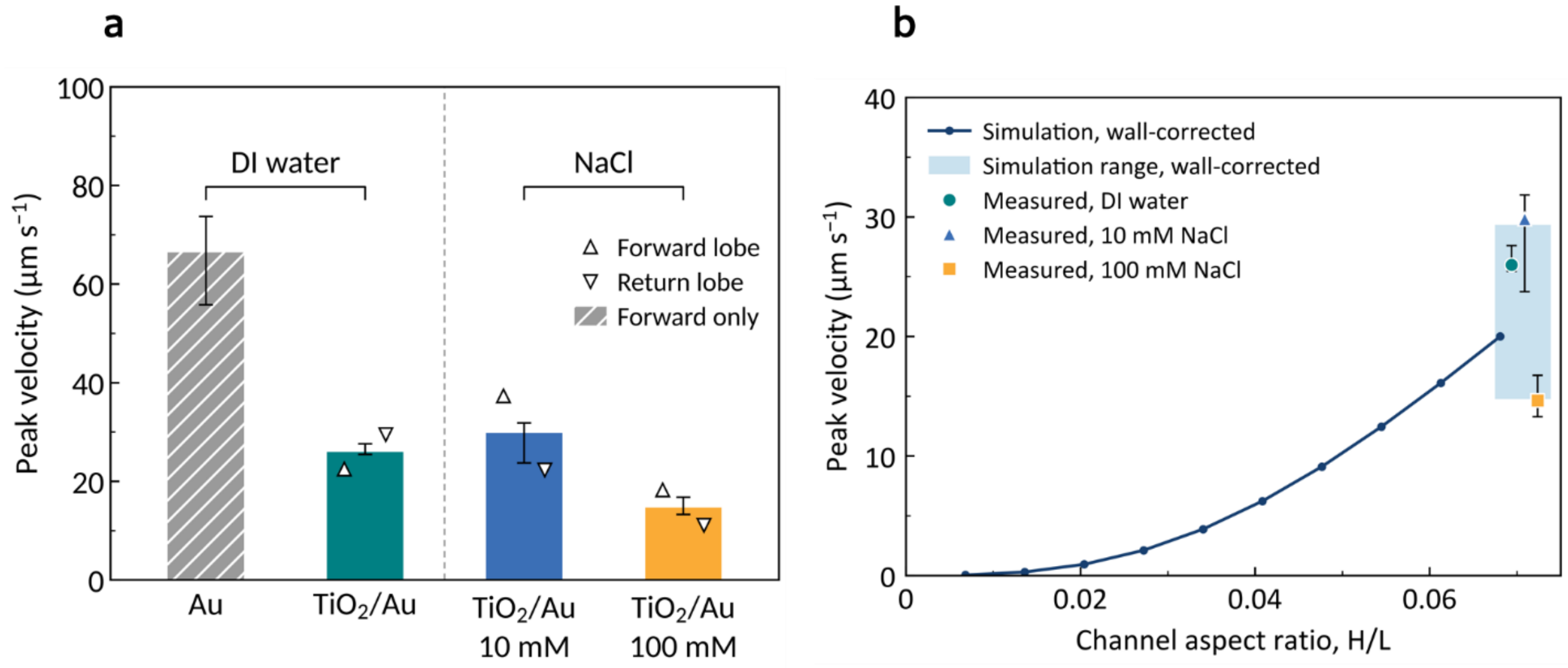

Figure 5. **| Flow peak velocity is independent of ionic strength but depends on channel geometry**. (a) Peak velocity of the $TiO_2$/Au membrane in deionized water and 10 and 100 mM NaCl. The Hadley profile is antisymmetric about mid-depth and has two lobes of opposite sign: a forward lobe in the upper half of the channel, where warm fluid flows from the hot well to the cold well, and a return lobe in the lower half, where cooler fluid flows back toward the hot well. The peak velocity (bar) is the mean magnitude of the two lobe maxima; symbols show the forward ( ▵ ) and return ( ▿ ) peaks. For the bare Au membrane (hatched), which has no return lobe, only the forward peak is shown, for reference; because its profile is not a closed Hadley roll, its speed is not compared with the $TiO_2$/Au runs. Error bars are bootstrapped 95% confidence intervals. Each run was performed on a different day after thermal equilibration (Supplementary Note 8). (b) Wall-corrected simulated peak velocity versus channel aspect ratio H/L, where H is the channel height and L is the 23.5-mm distance between wells (line). The shaded band spans the bounding thermal conditions for the current device (H/L = 0.068; absorbed heat flux, 490–530 W $m^{-2}$; $h_{top}$, 8–15 W $m^{-2}$ $K^{-1}$; $h_{bot}$, 5–10 W $m^{-2}$ $K^{-1}$; Methods). Measured peak velocities from a are overlaid as symbols, offset horizontally for clarity; error bars indicate bootstrapped 95% confidence intervals.

A membrane coated with gold alone, without $TiO_2$, also drives flow under illumination, but its flow runs in one direction only, which is inconsistent with a closed Hadley roll; we therefore treat it as a distinct flow regime and do not compare its speed (Supplementary Note 7). Notably, the Rayleigh number ($Ra_H$) scales with the fourth power of the channel height, making geometry a materials-independent lever for controlling flow strength. Simulations at fixed absorbed heat flux show that peak flow speed decreases sharply as the aspect ratio H/L decreases, falling below 1 μm $s^{-1}$ for $H/L < 0.02$ (Fig. 5b). The current cell ($H/L = 0.068$) is in the regime where photothermal convection dominates.

## DISCUSSION

Observing flow near an illuminated photocatalytic surface does not necessarily indicate photochemical activity or show that the membrane pumps liquid. In the $TiO_2$–Au membrane studied here, light is mainly absorbed by the gold layer. This absorption creates photothermal heating, which in turn induces a lateral (x-direction) temperature gradient that drives a Hadley roll on each side of the membrane. Any net flow across the membrane is below our measurement resolution (about 2 μm $s^{-1}$).

Materials and optical properties determine whether light induces chemical reactions or generates heat. At 369 nm, the gold film absorbs most of the incident light, while the thin $TiO_2$ overlayer reduces reflection and enhances heat generation in the gold. Titania sputtered at room temperature onto polycarbonate is expected to be amorphous[12,13] rather than the photocatalytically active anatase (which would require annealing the apparatus at a temperature well above the glass transition temperature of polycarbonate, which would destroy the pores). As a result, a coating intended to render a surface photoactive may instead act primarily as an antireflection layer that enhances photothermal heating in the underlying metal. Previous studies on $TiO_2$–Au Janus wall-mounted pumps attributed flow to photoelectrokinetic mechanisms[7]. Our results do not contradict these findings but demonstrate that at the millimeter scale, with a strongly absorbing metal, thermal effects can dominate. Routine calculation of the absorption in the actual layer stack is therefore essential before attributing observed flows to photochemistry.

Device design critically influences whether heat can drive fluid motion. The Rayleigh number ($Ra_H$) scales with the fourth power of channel height; thus, the same absorbed power that generates flow in a 1.6 mm cell produces negligible motion in channels with H/L below approximately 0.02. Additional factors, including side walls, slide thermal conductivity, and environmental conditions, are equally significant. For example, sapphire slides nearly eliminate the temperature gradient, while simulations predict a weaker gradient for the same cell on a temperature-controlled stage than on a microscope stage in still air. Consequently, flow speed depends on often unreported variables such as substrate material, mounting, and ambient conditions. Nonetheless, the shape of the velocity profile is more robust: a closed, reversing cubic profile consistently indicates buoyant recirculation, regardless of flow magnitude. Therefore, comparisons of flow speeds across different salts or materials are only meaningful when thermal environments are carefully matched.

Time is also a critical factor. Photothermal flows are governed by the heat balance of the entire cell, which in this study required about nine minutes to stabilize, much longer than the fluid's rapid response to temperature changes. Our velocity scans were performed 10–25 minutes after illumination, capturing steady-state flow. Measurements taken too early would under-report flow speed and could reveal asymmetries in the roll structure. In contrast, photoelectrokinetic flows originate at the active surface and are not bound to the cell's overall thermal equilibration. Monitoring the evolution and decay of flow upon switching the light on and off is therefore a straightforward way to distinguish between photothermal and photoelectrokinetic mechanisms.

These findings provide a practical framework for distinguishing photoelectrokinetic flows from photothermal convection: quantify absorption in the full material stack, map temperature, measure velocity across the channel depth, compare ionic-strength conditions after thermal equilibration, and test how geometry and substrate conductivity affect buoyancy. Although our two-dimensional model (with side-wall corrections) and measurements in a single device geometry constrain the scope of our conclusions, they point to thin channels, conductive substrates and weakly absorbing electrodes as strategies for isolating photoelectrokinetic effects. This distinction is important wherever light drives fluid motion—from convection-enhanced analyte delivery in surface biosensors[14,15], to particle patterning on illuminated electrodes[16,17], light-actuated microfluidic valves and pumps[18,19], and photocatalytic water treatment, where apparent photochemical gains may instead reflect photothermal mixing[20]. Identifying the dominant mechanism is therefore essential to optimizing these systems and interpreting their performance.

# METHODS

## Design and fabrication of the Janus Membrane

Track-etched polycarbonate (PC) membranes (Cyclopore, Whatman; 25 mm diameter, 12 μm pore diameter, 20 μm thickness) served as substrates. Porosity, determined from scanning electron microscopy (SEM) images analyzed in Fiji, was 7.3% (Supplementary Information). Before film deposition, membranes were sonicated in isopropyl alcohol (IPA) for 10 min, dried under $N_2$, and baked in a vacuum oven at 80 °C for 8 h to remove residual solvent and adsorbed moisture. This temperature is well below the PC glass transition (~147 °C) and prevents thermal deformation of pores. Thin films were deposited by sputtering (SD-400, PVD Products) at the Birck Nanotechnology Center, Purdue University. Gold (Au) was deposited on both faces of the membrane by DC sputtering at 50 W, with an Ar flow of 15 sccm and a chamber pressure of 5 mTorr, without an adhesion layer. Each face was coated for 10 min at 5 nm $min^{-1}$, yielding a nominal Au thickness of 50 nm. Titanium dioxide ($TiO_2$) was then deposited over the Au on one face by RF sputtering at 100 W under the same Ar flow and pressure, for 7.5 min at 2 nm $min^{-1}$, yielding a nominal thickness of 15 nm. The resulting Janus membrane structure was $TiO_2$ (15 nm)/Au (50 nm)/PC/Au (50 nm).

## Device Assembly

Each layer was laser-cut from a 1.6 mm PDMS sheet into a dumbbell profile (center-to-center well distance 23.5 mm) as described in the Results. The bottom layer was mounted on a 1-mm-thick fused-quartz slide and covered with 0.1 mm double-sided acrylic adhesive tape, sealing the observation channel from above while leaving both wells open. The membrane was then aligned over the hot well with its $TiO_2$ face oriented toward the LED. The assembly was completed by placing the top PDMS layer and a second quartz slide, with the cold wells aligned to form a continuous through-well.

## Flow Measurements

Tracer suspensions were prepared by adding one drop of 2 μm fluorescent polystyrene microspheres (Fluoro-Max G0200, Thermo Scientific; 1% solids) to 50 mL of deionized water or NaCl solution. Suspensions were sonicated for at least 5 minutes, then loaded into both layers with a micropipette. The UV LED (UV Beast V3 mini; nominal 365 nm, measured peak 369 nm, FWHM 10.5 nm) was operated at a constant 3.6 V (HM310, Hanmatek) and positioned about 5 mm above the top quartz slide over the hot well. Irradiance was measured using a radiometer (PMA2100, Solar Light) and ranged from 60 to 65 mW $cm^{-2}$ (mean 62.5 mW $cm^{-2}$). Minor variations were due to small differences in LED height between sessions. Each condition was initially scanned in the dark. The LED was then placed over the cell, and illuminated scans began 10 to 25 minutes later after the cell settled (simulated settling time about 9 minutes; Supplementary

Fig. S7). The LED remained on throughout. Each scan of 13 to 25 focal planes lasted 16 to 30 minutes. For velocity measurements, the cell rested on the microscope stage in still room air at about 20 °C. Images were acquired with an inverted microscope (Eclipse TE2000-U, Nikon) using a Plan Fluor 10× objective (NA 0.30) and a cooled CCD camera (pco.1600, PCO). Acquisition settings were 1400 × 1000 pixels, 0.74 µm per pixel, 50 ms exposure, and 500 ms delay. The frame interval from the camera's embedded timestamp was 0.6748 s for all sessions. Each plane was imaged for 42 to 58 frames.

## Height Calibration

Imaging was performed with a dry objective; thus, moving the stage by $\Delta z_s$ shifted the focal plane in water by $n_w \times \Delta z_s$, where $n_w = 1.33$ is the refractive index of water [21]. A 100-µm stage movement thus corresponded to a 133-µm shift in the focal plane. Both channel walls were located directly in each image stack. The floor was defined as the focal plane where particles adhered to the quartz appeared sharp, and the ceiling as the tape surface, visible as a particle-free band. Since focal planes were separated by 133 µm, the tape surface was placed at the midpoint between the last plane with moving particles and the first particle-free plane, giving a positional accuracy of ±67 µm. Depth z was measured downward from the ceiling (z = 0) to the floor (z = −H = −1600 µm; Fig. 2b) and calculated as

$$z = -H + n_w(z_s - z_{s,0}),$$

where $z_{s,0}$ is the stage position focused on the floor and H = 1.6 mm is the channel height. Depths were normalized as

$$Z^* = \frac{z}{H} + \frac{1}{2},$$

so that Z* runs from −0.5 at the floor to +0.5 at the ceiling, and Z*H is the height above mid-depth used in Hadley theory. For the 100 mM run, the floor was not imaged, so depths were referenced from the ceiling,

$$z = n_w(z_s - z_{s,c}),$$

where $z_{s,c}$ is the stage position focused on the ceiling.

## Particle tracking

Particles were detected and linked into trajectories using TrackPy 0.7[22], which implements the Crocker–Grier algorithm[23]. Identification parameters included a feature diameter of 15 px, minimum integrated brightness of 800, minimum separation of 15 px, search range of 90 px per frame, one-frame memory, and a minimum track length of 8 frames. These settings were applied uniformly to all planes in every experiment. Image sequences were acquired at single, known focal planes; therefore, particle depth was assigned according to the focal position. Out-of-focus particles were excluded by applying a mean peak brightness threshold, determined from control planes inside the tape where no particles could be in focus. Among 14 such planes, the brightest track reached 155 counts above background; thus, only tracks exceeding 200 counts were retained. No tracks from tape planes passed this threshold, ensuring that particles one focal step (≈133 µm) away from the imaged plane were excluded. The streamwise velocity of each track was calculated as the median frame-to-frame displacement multiplied by the pixel size (0.74 µm) and divided by the frame interval (0.6748 s). For each plane, the median velocity across all tracks was reported, with a 95% confidence interval estimated from 10,000 bootstrap resamples. The median was used due to low track counts in several planes. Results are shown for planes with at least three tracks. Two controls validate the measurements: particles adhered to the quartz floor showed median velocities of 0.1 μm $s^{-1}$ in DI water and 0.002 μm $s^{-1}$ in 10 mM NaCl, consistent with a no-slip condition.

## Control Experiments

Three controls were run with the same device, optics and tracking settings. (i) Dark runs: each condition was recorded at every focal plane without UV illumination, and the median velocity was indistinguishable from zero at all planes (Supplementary Fig. S9). (ii) Bare-Au membrane: a polycarbonate membrane coated with 50 nm Au on both faces and no $TiO_2$ was illuminated under the same conditions. (iii) Ionic strength: $TiO_2$/Au membranes were tested in deionized water and in 10 mM and 100 mM NaCl, and one membrane was tested at both 10 mM and 100 mM. Salt runs were recorded on different days, with irradiance between 60 and 65 mW $cm^{-2}$ and in still room air, so differences in heat loss between days are expected to change the flow speed (Supplementary Notes 6 and 8).

## Statistics

Each plane is summarized by the median streamwise velocity of its in-focus tracks, with a 95% confidence interval from 10,000 bootstrap resamples of the tracks. Planes with fewer than three tracks are not shown. Rayleigh numbers were obtained by least-squares fits of the wall-corrected Hadley profile to the plane medians, and $R^2$ is reported for each fit.

## Optical Absorption

The fraction of incident light absorbed in the membrane coating was determined using a coherent transfer-matrix method[24] for the planar stack: water | $TiO_2$ (15 nm) | Au (50 nm) | polycarbonate, at normal incidence and at the LED peak wavelength ($\lambda$ = 369 nm). Because the fused-quartz

window and water layer are transparent at 369 nm and much thicker than the LED coherence length ($\ell_c \approx 13$ μm, $\Delta\lambda = 10.5$ nm), optical propagation through them is incoherent. Water and polycarbonate were therefore modeled as semi-infinite incident and exit media, respectively, neglecting upstream reflection losses (~4%) at the air–quartz–water interfaces. The $TiO_2$ and Au films, much thinner than $\ell_c$, were treated coherently using nominal deposition thicknesses. Complex refractive indices ($\tilde{n} = n + i\kappa$) at 369 nm were taken as: water[21], 1.347; $TiO_2$[25], 2.712 + 0.010$i$ ; Au[9], 1.478 + 1.898$i$ ; polycarbonate[26], 1.638. At normal incidence, Fresnel reflection and transmission coefficients across interface $jk$ were calculated as:

$$r_{jk} = \frac{\tilde{n}_j - \tilde{n}_k}{\tilde{n}_j + \tilde{n}_k},\ t_{jk} = \frac{2\tilde{n}_j}{\tilde{n}_j + \tilde{n}_k},$$

with phase accumulation $\delta_j = 2\pi\tilde{n}_j d_j/\lambda$ across each film of thickness $d_j$. Multiplying the interface and propagation matrices across the stack accounts for all coherent multiple reflections. To partition the total stack absorptance ( $A_{\text{total}} = 1 - R - T$ ) into individual film contributions, the absorptance $A_j$ of each layer was obtained from the drop in the time-averaged Poynting flux $S(z)$ across its local thickness ( $0 \leq z \leq d_j$ ), normalized to the incident flux $S_{\text{inc}}$ :

$$A_j = \frac{S_j(0) - S_j(d_j)}{S_{\text{inc}}}$$

The stack yielded a reflectance of $R = 0.039$ and transmittance of $T = 0.074$ into the polycarbonate substrate. The Au and $TiO_2$ films absorbed $A_{Au} = 0.874$ and $A_{TiO_2} = 0.012$, respectively, with gold accounting for ~99% of the absorbed power ($R + A_{TiO_2} + A_{Au} + T = 0.999 \approx 1$). Residual absorption of the transmitted fraction ($T = 0.074$) within the polycarbonate bulk and Au film was neglected, providing a conservative bound on membrane heating. Assuming complete photothermal conversion and treating the perforated surface via geometric area scaling (excluding open pores of area fraction $\phi = 0.073$), the effective heat flux applied in simulations was calculated as:

$$q'' = (1 - \phi)\left(A_{\mathrm{TiO_2}} + A_{\mathrm{Au}}\right)I_0$$

This flux was applied as a boundary heat source on the membrane surface, which lies between the upper and lower water layers. Its partition between the two layers was not prescribed; it follows from the coupled heat-transfer solution and is set by the thermal resistance of each path to the surroundings (water layer, quartz slide and external heat-transfer coefficient above and below the membrane). For the measured incident irradiance of $I_0$ = 60–65 mW $cm^{-2}$ (see Flow Measurements), this gives q″ = 493-534 W $m^{-2}$. Consequently, simulations were parameterized using 490, 510, and 530 W $m^{-2}$ to represent the lower bound, mean, and upper bound, respectively.

## Numerical Model

Coupled flow and heat transfer were simulated in COMSOL Multiphysics 6.0[27] using the Laminar Flow and Heat Transfer in Solids and Fluids interfaces connected via Nonisothermal Flow coupling. The computational domain comprised a two-dimensional vertical cross-section of the cell along the channel axis (Supplementary Fig. S5), extending from the center of the hot well ($x = 0$) to the center of the cold well ($x = 23.5$ mm), with the observation channel spanning $9.5 \leq x \leq 20.5$ mm. Two water layers of height H = 1.6 mm were separated by a 0.1 mm solid partition and bounded by 1 mm fused-quartz slides. Over the hot well ($0 \leq x \leq 9.5$ mm), the partition represented the membrane and was modeled as an impermeable polycarbonate solid; beyond it ($9.5 \leq x \leq 20.5$ mm), it represented the adhesive tape and was modeled as acrylic. The two water layers therefore connected only through the cold well ($20.5 \leq x \leq 23.5$ mm).

Density was assumed constant ($\rho_0 = \rho(T_0)$ with $T_0 = 20\,°\mathrm{C}$) throughout, except in the buoyancy term, where the full temperature-dependent density of water was used. This approach corresponds to the Boussinesq approximation without linearizing $\rho(T)$. The steady-state velocity $\mathbf{u}$, pressure $p$ (referenced to the hydrostatic datum at $\rho_0$), and temperature $T$ in the water satisfy

$$\nabla \cdot \mathbf{u} = 0,$$

$$\rho_0(\mathbf{u} \cdot \nabla)\mathbf{u} = -\nabla p + \nabla \cdot [\mu(T)(\nabla\mathbf{u} + \nabla\mathbf{u}^{\mathrm{T}})] + [\rho(T) - \rho_0]\mathbf{g},$$

$$\rho_0 c_p(T)\,\mathbf{u} \cdot \nabla T = \nabla \cdot [k(T)\nabla T],$$

Here, $\mathbf{g} = -g\mathbf{e}_z$ is directed downward, so warmer, less dense water is subject to an upward buoyant force. Material properties were sourced from the COMSOL 6.0 material library[27]: temperature-dependent values for water (Supplementary Table S3), $k_s = 1.4\ \mathrm{W\,m^{-1}\,K^{-1}}$ for fused quartz[27,28], a temperature-dependent conductivity of approximately 0.2 W $\mathrm{m^{-1}}$ $\mathrm{K^{-1}}$ for polycarbonate (membrane), and a conductivity of approximately 0.18 W $\mathrm{m^{-1}}$ $\mathrm{K^{-1}}$ for acrylic[27] (tape). Heat conduction in solids was modeled as

$$\nabla \cdot (k_s \nabla T) = 0.$$

Light absorbed in the coating was modeled as a boundary heat source, $q''$ (see Optical absorption), on the illuminated membrane face over $0 \leq x \leq 9.5$ mm. All solid–fluid interfaces were assigned no-slip boundary conditions. The vertical end planes at $x = 0$ and 23.5 mm were thermally insulated and treated as symmetry boundaries. Pressure was fixed at a single reference point, as the cell is closed. The outer face of the quartz slides lost heat according to

$$-\mathbf{n} \cdot \mathbf{q} = h(T - T_\infty).$$

The two experimental sessions, each with distinct thermal environments, were modeled separately (Supplementary Table S4). During infrared imaging, the cell was placed on a temperature-controlled stage at 22 °C. The top face lost heat to air at 22 °C ($h_{\text{top}} = 8\text{–}15\ \text{W}\,\text{m}^{-2}\,\text{K}^{-1}$), while the contact resistance between the bottom slide and the stage (not directly measured) was represented by a conductance $h_c = 100\ \text{W}\,\text{m}^{-2}\,\text{K}^{-1}$ to the stage at 22 °C (equivalent to an air gap of about 0.26 mm) in all infrared-session simulations . During velocity measurements, the cell rested on the microscope stage in still air at 20 °C, so both faces lost heat via natural convection ($h_{\text{top}} = 8\text{–}15$ and $h_{\text{bot}} = 5\text{–}10\ \text{W}\,\text{m}^{-2}\,\text{K}^{-1}$, nominally 12 and 8 $\text{W}\,\text{m}^{-2}\,\text{K}^{-1}$, respectively). For each session, the upper and lower bounds of the simulated range combined the largest heat input with the weakest cooling, and the smallest heat input with the strongest cooling.

The equations were solved using a fully coupled Newton method and the PARDISO direct solver, with a relative tolerance of $10^{-6}$. The structured quadrilateral mesh was refined by halving the element size twice; between the two finest meshes, changes in peak speed, well temperatures and fitted Rayleigh number were less than 1% (final mesh: 51,600 domain and 2,408 boundary elements; Supplementary Table S5). The global energy balance was satisfied to better than $10^{-6}$. Time-dependent simulations under velocity-session conditions indicated that the peak channel velocity reached 99% of its steady-state value approximately 9 minutes after illumination (Supplementary Fig. S7).

Simulated temperatures were extracted along the outer surface of the top slide from $x = 0$ to 23.5 mm, matching the line sampled in infrared imaging. Velocity profiles $u(z)$ were obtained across the bottom water layer at $x = 14$, 15, and 16 mm, corresponding to the particle tracking field of view within 1 mm of the channel center. Each profile was fitted with the Hadley profile to determine the local temperature gradient and Rayleigh number (see Hadley theory). As the 2D model omits the side walls of the 1 mm-wide channel ($W/H = 0.625$), each simulated profile was converted to the 3D channel centerline using the analytical four-wall Hadley solution at the same axial temperature gradient. This procedure reduced the peak speed by a factor of 0.73 without affecting $\mathrm{Ra}_H$, as heat transfer was dominated by conduction (Péclet number $\approx 0.3$).

## IR Thermometry

Temperature was measured from above using a cooled InSb camera (QFI InfraScope, MWIR-1024; 3–5 µm; 1/6× MWIR objective; 80 Hz). The device was placed on a temperature-controlled stage (QFI TSFA-4237) set to 22 °C and imaged at normal incidence through a 1 mm GE 124 fused-quartz top slide. The infrared signal represents an average over the depth from which the camera collects emission, primarily within the quartz slide and the water directly beneath it. Thermal simulations show that the temperature varies by no more than 0.25 K across the 1 mm slide; therefore, the slide does not significantly bias the inferred temperature of the upper water layer. For comparison with the measurements, simulated temperatures were taken at the outer surface of the top slide. In the camera software, the emissivity of the imaged surface (quartz slide backed by water) was set to 0.95, and the reflected temperature was set to 22.0 °C. In this spectral range, quartz backed by water reflects 2–3%[29] , corresponding to an effective emissivity of 0.976. The software setting therefore slightly overestimates temperature rises, less than 3% (about 0.1 K here). Before illumination, the image showed a uniform temperature of 21.98 ± 0.03 °C. The membrane was illuminated at 60–65 mW cm$^{-2}$, and images were acquired after 10 min, once the temperature had stabilized. The temperature profile along the 23.5 mm line from the hot-well center (P1) to the cold-well center (P2)—the section modeled in the simulations—was sampled by cubic-spline interpolation[30]. Separately, thermocouples placed in the upper water layer at the centers of the two wells and connected to a dual-input thermometer (Amprobe TMD-56) recorded temperatures of 25.4 °C and 21.9 °C after 10 min of illumination. Thermocouples provided the absolute temperatures, while the infrared image provided the spatial distribution.

## Hadley Theory

In a long, shallow layer ($H \ll L$) subjected to a locally uniform horizontal temperature gradient ($\gamma = \partial T/\partial x$), the flow far from the end walls is parallel to the boundaries. Inertia is negligible ($Re \ll 1$), and buoyancy is treated using the Boussinesq approximation. Measuring z here from mid-depth,

$$z = Z^{*}H,$$

with no slip at the top and bottom walls ($z = \pm H/2$) and no net flux through the layer, the resulting velocity follows the cubic Hadley profile[10,11].

$$u_{2D}(z) = \frac{g\beta\gamma}{6\nu}\left(z^3 - \frac{H^2}{4}z\right),$$

whose amplitude is set by the horizontal Rayleigh number

$$\mathrm{Ra}_H = \frac{g\beta\,|\gamma|\,H^4}{\nu\kappa}.$$

In a channel of finite width W with no-slip side walls at $y = \pm W/2$, the profile is reduced near the side walls. Adding to $u_{2D}$ a Fourier series in z that cancels it at the side walls gives

$$u_{3D}(y,z) = u_{2D}(z) - \frac{g\beta\gamma H^3}{4\pi^3\nu}\sum_{n=1}^{\infty}\frac{(-1)^n}{n^3}\,\frac{\cosh(2n\pi y/H)}{\cosh(n\pi W/H)}\,\sin\left(\frac{2n\pi z}{H}\right),$$

which at the centerline ($y = 0$) lowers the peak speed by a factor of 0.73 for $W/H = 0.625$ (Supplementary Note 4). Measured velocities were fitted with $u_{3D}(0, z)$ by least squares, with $\gamma$ as the only free parameter, using water properties at 25 °C, representative of the mean fluid temperature (CoolProp)[31]. Simulated profiles were fitted with $u_{2D}$ to obtain their local $\gamma$ and then converted to the channel centerline with $u_{3D}$ at the same $\gamma$. The measured profiles show no net flow greater than approximately 2 μm $s^{-1}$. Moreover, because a net flow is symmetric about the mid-height while the Hadley profile is antisymmetric, a small net flow has little effect on the fitted $\gamma$.

## Use of AI tools

Claude (Anthropic), accessed via claude.ai from June to November 2026, was used during data analysis and manuscript preparation in both chat and code formats. During the analysis, it assisted with editing text for clarity and debugging Python scripts for the particle-tracking pipeline, transfer-matrix calculations, and velocity-profile fitting. It also helped write code to generate figures from raw data files. Claude Code identified inconsistencies in the data and records, including the camera frame interval in image timestamps, the recording times of each focal plane relative to dark scans, minimum track counts per plane, and differences in thermal conditions between the infrared-imaging and microscope stages. In addition, it was used to verify mathematical derivations and suggest edits. The authors checked all flagged issues against raw images, instrument records, and laboratory notes before making any changes. The authors made all analysis decisions and interpretations and reached all conclusions. They reviewed all code and text and take full responsibility for the content.

## Supplementary Information

The Supplementary Notes, Figures and Tables cited in the text will be provided in a later version of this preprint.

## Data and code availability

Data and COMSOL model files are available from the corresponding author upon reasonable request and will be deposited in Zenodo upon publication.

## Competing interests

The authors declare no competing interests.

## Acknowledgements

We thank Zhengwei Chen for initial suggestions on particle tracking velocimetry and Sohum Sodhi for laser cutting the PDMS sheets. Thin-film deposition was carried out at the Birck Nanotechnology Center, Purdue University. This work was partially supported by the Ralph W. and Grace M. Showalter Research Trust. S.A. acknowledges support from the Lambert Teaching Fellowship of the School of Mechanical Engineering, Purdue University, and the Bilsland Dissertation Fellowship.

## Author contributions

S.A. conceived the study, designed and fabricated the membranes and devices, carried out the particle-tracking, infrared and thermocouple measurements, developed the analysis and the simulations, and wrote the manuscript. Y.F. conceived the study, designed and fabricated the membranes and devices, carried out the particle-tracking. L.F.: carried out particle-tracking analysis; fabricated device; helped with figures. M.I.: carried out particle tracking experiments; helped with the figures. S.S.: led the infrared thermometry measurements. T.H.: performed multiphysics simulations. J.N.: Writing. S.R.: Characterization of membranes S.W. advised on particle-tracking velocimetry and provided microscope facility. A.M. provided the infrared thermometry facility and advised on the measurements. T.B. advised on the optical modeling. J.L.M. supervised the project and edited the manuscript. D.M.W. supervised the project, acquired funding and edited the manuscript. All authors reviewed and approved the manuscript.